\documentclass[journal]{IEEEtran}

\usepackage{mflogo,texnames}
\usepackage{enumerate}
\usepackage{float}
\usepackage{hyperref}       
\usepackage{url}            
\usepackage{booktabs}       
\usepackage{amsfonts}       
\usepackage{nicefrac}       
\usepackage{microtype}      
\usepackage{times}
\usepackage{xcolor}
\usepackage{soul}

\usepackage{lineno,hyperref}
\usepackage{enumerate}
\usepackage{amsfonts}       
\usepackage{graphicx}  
\usepackage{subfigure}
\usepackage{multirow}
\usepackage{amsmath}
\usepackage{float}
\usepackage{color}
\usepackage[ruled]{algorithm2e}

\ifCLASSINFOpdf
\else
\fi
\begin{document}
%
\title{TEGDetector: A Phishing Detector that Knows Evolving Transaction Behaviors }
%
%
%

\author{Jinyin Chen,
	Haiyang Xiong,
	Dunjie Zhang,
	Zhenguang Liu,
	and Jiajing Wu
}

\maketitle



\begin{abstract}
	Recently, phishing scams have posed a significant threat to blockchains. Phishing detectors direct their efforts in hunting phishing addresses. Most of the detectors extract target addresses' transaction behavior features by random walking or constructing static subgraphs. The random walking methods, unfortunately, usually miss structural information due to limited sampling sequence length, while the static subgraph methods tend to ignore temporal features lying in the evolving transaction behaviors. More importantly, their performance undergoes severe degradation when the malicious users \emph{intentionally} hide phishing behaviors. To address these challenges, we propose \emph{TEGDetector}, a dynamic graph classifier that learns the evolving behavior features from transaction evolution graphs(TEGs). \emph{First}, we cast the transaction series into multiple time slices, capturing the target address's transaction behaviors in different periods. \emph{Then}, we provide a fast non-parametric phishing detector to narrow down the search space of suspicious addresses. \emph{Finally}, TEGDetector considers both the spatial and temporal evolutions towards a complete characterization of the evolving transaction behaviors. Moreover, TEGDetector utilizes adaptively learnt time coefficient to pay distinct attention to different periods, which provides several novel insights. Extensive experiments on the large-scale Ethereum transaction dataset demonstrate that the proposed method achieves state-of-the-art detection performance. 
\end{abstract}


\section{Introduction}
\noindent As a decentralized and distributed public ledger, blockchain technology~\cite{2017The} has enjoyed great success in various fields, \emph{e.g.}, finance, technology, and culture~\cite{2017The}. Cryptocurrency~\cite{wood2014ethereum,yuan2018blockchain}, undoubtedly, is one of the most profound applications of blockchain. As the largest blockchain platform supporting smart contracts, Ethereum now holds cryptocurrencies worth more than $\$39.3$ billion dollars. Unfortunately, the decentralization of blockchain also breeds numerous financial scams~\cite{vasek2015there, chen2018detecting,vasek2018analyzing,huang2020understanding}.  Chainalysis\footnote{A provider of investigation and risk management software for virtual currencies} has reported that phishing scams, which accounted for 38.7\% of all Ethereum scams~\cite{EtherScamDB}, stole $\$ $34 million from Ethereum platform in 2018.  Phishing  refers to the  impersonating a website of an honest firm, which obtains the users sensitive information and money via phishing websites. Recently, phishing scams are reported every year, and they become even more sophisticated.


As a result, detecting phishing addresses for blockchain has attracted widespread attention. Fundamentally, phishing detection aims to learn a mapping function that bridges their historical transaction behaviors to a binary output $y$, where $y=0$ denotes the target address is normal and $y=1$ represents it is abnormal. A family of works~\cite{li2020identifying,lin2019evaluation} utilized manually extracted features to capture the target user's transaction behaviors. Unfortunately, these models require an incisive data understanding, leading to unsatisfactory results.

Recent approaches explored using different sorts of graph embedding algorithms to address the issue. \emph{Walking based detectors}~\cite{perozzi2014deepwalk,grover2016node2vec,wu2020phishers,lin2020t,yuan2020phishing} direct their efforts at adopting random walking to characterize the temporal evolution between transactions. \emph{Subgraph based detectors}~\cite{shen2021identity,wang2021tsgn,yuan2020phishing,zhang2021mcgc} usually described the target address's transaction pattern through a static subgraph. Specifically, they first construct the transactions of the target address and its neighbors in all periods into a static subgraph, then built upon the success of graph neural networks to learn the spatial graph structure from the static subgraphs. All in all, there are several challenges in this task.

\textbf{Balance between structural and temporal information.} Earlier works tend to focus on either structural or temporal information, which leads to considerable information loss. This motivates us to consider whether we can combine and balance them to approach better detection performance. Due to the incompleteness of the structural information caused by the limited sampling sequence length, it is difficult for the walking strategy to achieve such a balance. We speculate that one viable approach is to construct multiple transaction subgraphs for a target address, where each subgraph characterizes the transaction topology within a temporal period.  We term the transaction subgraphs as \emph{dynamic subgraphs}. Taking the Ethereum~\cite{wood2014ethereum} users as an example. We apply the static subgraph construction method in MCGC~\cite{zhang2021mcgc} and extend it to the dynamic subgraph construction. 


\textbf{Robustness against hidden phishing addresses.} Researches~\cite{zugner2018adversarial,dai2018adversarial,chang2020restricted} on the vulnerability of the graph analysis methods also reveal potential security issues in blockchain phishing detection. Intuitively, the phishers may bypass the detection by transacting to specific addresses. To verify the robustness of existing phishing detectors, we randomly add transactions between the first and second-order neighbor addresses of 200 verified Ethereum phishing addresses. 

To address these challenges, our approach converts phishing detection into a dynamic graph classification problem. We construct a series of transaction evolution graphs (TEGs) for multiple time slices, which have the key advantage of retaining both the spatial structural and temporal information. To effectively utilize the abundant information contained in TEGs, we also propose TEGDetector that serves to capture target addresses' behavior features. The TEGDetecor is composed of graph convolutional layers and GRU, which respectively capture the topology structure and dynamic evolution characteristics of the network. Specifically, we introduce adaptive time coefficient to comprehensively balance the user's behavior features in all periods, rather than using only the one in the most recent period. This benefits exploring the crucial factors of phishing detection and helps TEGDetector identify possible malicious deception. Our contributions may be summarized as follows:

\begin{itemize}
	\item To the best of our knowledge, this is the first work that defines the phishing detection task as a dynamic graph classification problem. The proposed method balances the structural and temporal information through the constructed TEGs, and provides TEGDetector to map these information into user behavior features.
	\item A fast non-parametric phishing detector (FD) is presented, which can quickly narrow down the search space of suspicious addresses and improve the detection performance and efficiency of the phishing detector.
	\item Experiments conducted on the Ethereum dataset demonstrate that TEGDetector can achieve state-of-the-art detection performance. Interestingly, phishing deception experiments caused the existing methods to undergo an accuracy decline of 25-50\%, while TEGDetector achieves more robust phishing detection with only a decline of 13\%.
	
	
\end{itemize}

The rest of the paper is organized as follows. Related works are introduced in Section \uppercase\expandafter{\romannumeral2}, while the proposed method is detailed in Section \uppercase\expandafter{\romannumeral3}. Experiment results and discussion are showed in Section \uppercase\expandafter{\romannumeral4}. Finally, we conclude our work.


\section{Related Work}
In this section, we briefly review the existing works on phishing detection and graph classification.
\subsection{Phishing Detector}
To provide early warnings to potential victims, various phishing detectors are proposed to identify phishing addresses. 

\textbf{Feature engineering based phishing detectors}~\cite{lin2019evaluation, li2020identifying} usually manually extract basic and additional transaction statistical features from preprocessed transactions, then use them to train a classifier. To realize automatic phishing behavior feature extraction, \textbf{walking based phishing detectors} learn the user's transaction behavior feature unsupervised. Wu et al.~\cite{wu2020phishers} performed a biased walking according to the transaction amount and timestamp, then obtained the address sequence to extract the user's behavior features. Lin et al.~\cite{lin2020t} further defined the temporal weighted multidigraph (TWMDG), which ensures the walking sequences contain the actual meaning of the currency flow. The \textbf{subgraph based phishing detectors} pay more attention to spatial structure information. Yuan et al.~\cite{yuan2020phishing} designed second-order subgraphs to represent the target address, modeling the phishing detection task as a graph classification problem. Wang et al.~\cite{wang2021tsgn} mapped the original transaction subgraphs to the more complex edge subgraphs. Shen et al.~\cite{shen2021identity} and Zhang et al.~\cite{zhang2021mcgc} introduced graph neural networks to realize blockchain phishing detection in an end-to-end manner.

In general, the existing blockchain phishing detectors will sacrifice some structural or temporal information when capturing users' behavior features. Moreover, the robustness of phishing detectors lacks research.

\subsection{Graph Classification}
For a blockchain transaction platform, the transactions of the target address and its neighbors are usually sufficient to reflect its transaction pattern. Intuitively, it is possible to convert the phishing detection task into a graph classification problem.

There are two general approaches to graph classification. The first~\cite{borgwardt2005shortest,shervashidze2011weisfeiler, vishwanathan2010graph,yanardag2015deep} assumes that molecules with similar structures share similar functions, and converts the core problem of graph learning to measure the similarity of different graphs. The second~\cite{DBLP:conf/nips/DefferrardBV16,DBLP:conf/aaai/ZhangCNC18,ying2018hierarchical,lee2019self} introduces various pooling operations to aggregate the node level representations into the graph level, which performs better on complex graphs.

It is worth noting that although numerous dynamic graph mining methods~\cite{DBLP:journals/corr/ChungGCB14, goyal2018graph,li2018deep,chen2019lstm,chen2019generative} have been studied, most graph classifiers are designed for static graphs. Due to the dynamic evolving pattern of user behaviors, a dynamic graph classifier will be beneficial to phishing detection.

%
%

\begin{figure*}[htb]\setlength{\abovecaptionskip}{-0.3cm}
	\centering
	\includegraphics[width=0.85\linewidth]{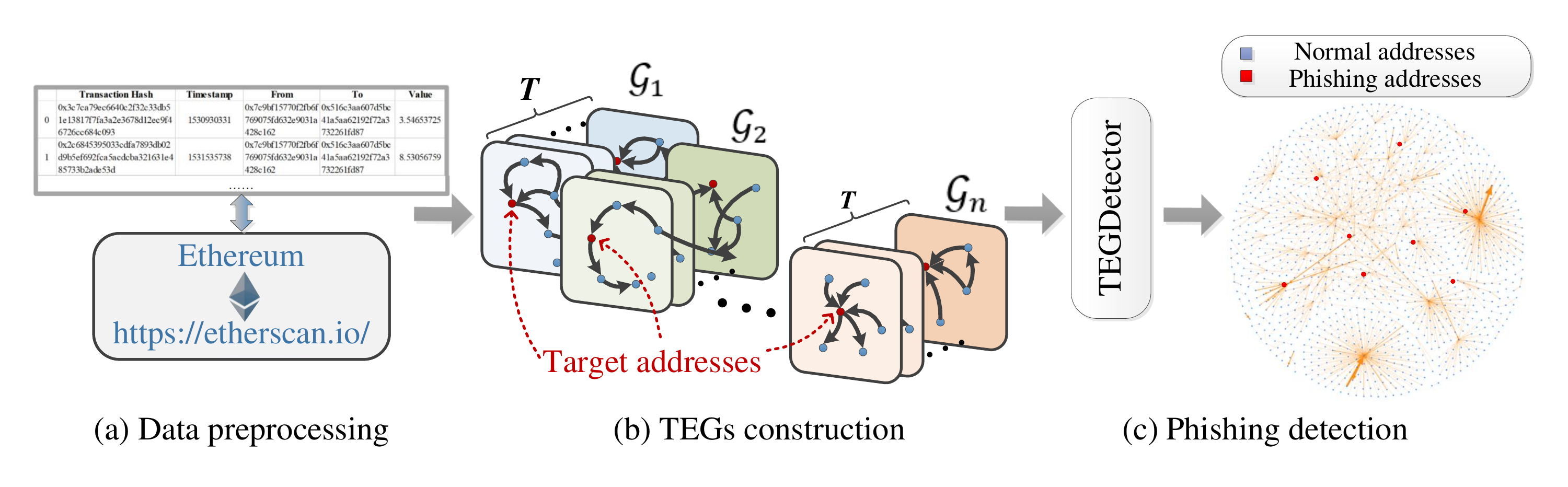}\\
	\caption{A high-level overview of our pipeline. (a) Preprocessing data from Ethereum. (b) Construction of TEGs for each address. (c) Phishing detection via TEGDetector.}
	\label{fig.flow}
\end{figure*}

\section{Methodology}

TEGDetector seeks to identify phising addresses by extracting their evolving behavior cues from the transaction graphs. An overview of the proposed method is outlined in Figure~\ref{fig.flow}. In the following,
we present the details of each component.

\subsection{Data Preprocessing }\label{sec3.1}
The phishing detection problem on the blockchain is a typical supervised learning problem, which requires labeled user addresses to train TEGDetector. Here, we obtained an Ethereum address list from the blockchain academic research data platform Xblock\footnote{http://xblock.pro/}.
We extract transaction sending/receiving addresses, transaction amount, timestamps, and address labels as the crucial information for constructing TEGs. The sending addresses and the receiving ones correspond to the nodes on graph, and the transaction amount and timestamps represent the edge weight and temporal information between the node pairs, respectively. Moreover, we construct the address label-based attribute $X\in \mathbb R^{N\times 2}$ for $N$ addresses. $X_{i,1}=1$ if the address $v_i$ is a phishing address, and $X_{i,0}=1$ otherwise.

\subsection{TEGDetector}

\begin{figure*}[htbp]\setlength{\belowcaptionskip}{-0.5cm}\vspace{-2em}\setlength{\abovecaptionskip}{0.1cm}
	\centering
	\includegraphics[width=0.8\linewidth]{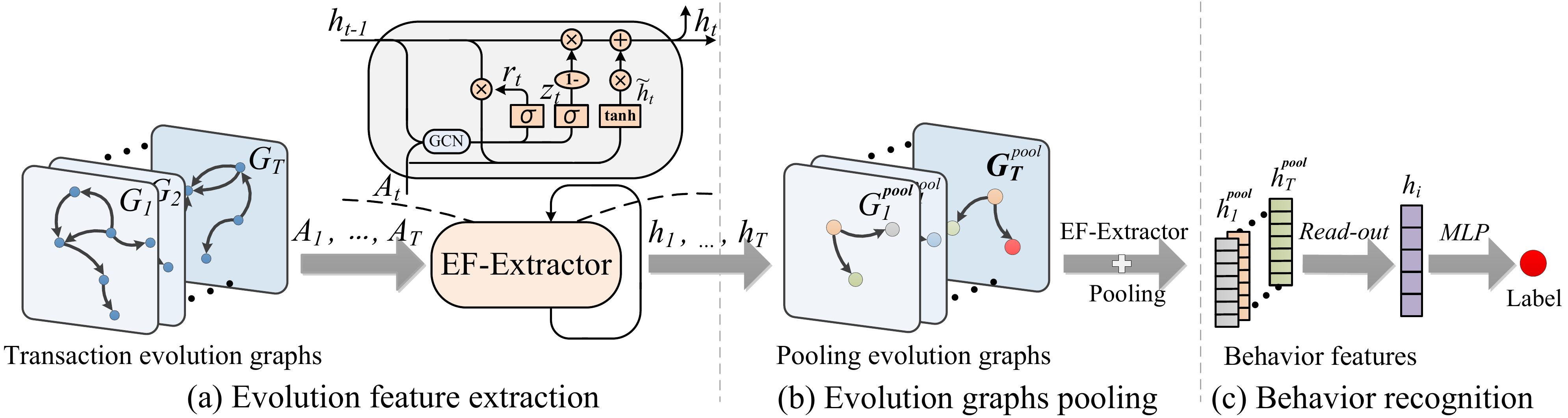}\\
	\caption{ The framework of TEGDetector. (a) EF-Extractor learns the evolution features of different TEG slices. (b) The evolution graphs pooling alternates with EF-Extractor to gradually aggregate the node-level evolution features into the graph-level. (c) The Read-out operation and Multilayer Perceptron (MLP) output the detection results.  }
	\label{fig4}
\end{figure*}
In this section, we designe a phishing detector for fully extracting the structure and temporal information from TEGs, termed TEGDetector. As shown in Figure~\ref{fig4}, TEGDetector is designed in an end-to-end manner, including evolution feature extraction (EF-Extractor), evolution graphs pooling, and behavior recognition.

The EF-Extractor integrates structural and temporal information to extract the addresses' evolution features.
Through the alternation with EF-Extractor, the evolution graph pooling aggregates the evolution features of similar addresses until obtaining the TEG's graph-level features. The behavior recognition assigns time coefficients to these graph-level features, and comprehensively considers the target address's transaction behaviors in different time slices, which also enhances the robustness of TEGDetector.

\textbf{EF-Extractor.} We introduce EF-Extractor to learn the user addresses' transaction evolution features at different time slices. Since graph convolutional layers have proven its powerful ability to capture the structural features of graphs in \cite{DBLP:conf/iclr/KipfW17} \cite{20K-Core}, EF-Extractor employs graph convolutional layers to learn the structural features of the current TEG slice. Meanwhile, we learn from the idea of GRU~\cite{DBLP:conf/emnlp/ChoMGBBSB14} to capture the temporal information of the TEGs. Another reason for choosing GRU is that it has fewer model parameters and runs faster than long short-term memory~\cite{97LSTM}.

Specifically, EF-Extractor utilizes a two-layer GCN~\cite{DBLP:conf/iclr/KipfW17} module to map the structural information to a $d$-dimensional node representation $Z$. As the structural features of the $t$-th slice of TEGs, $Z_t$ can be defined as:

\setlength{\parskip}{-0.4\baselineskip}
\begin{equation}
	\setlength{\abovedisplayskip}{0.2pt}
	\label{eq2}
	Z_t = GCN(h_{t-1},A_t)=f(\hat A_t \sigma(\hat A_t h_{t-1} W_0)W_1)
\end{equation}
where $\hat A_t = \tilde D_t ^{-\frac{1}{2}} \tilde A_t \tilde D_t ^{-\frac{1}{2}}$, $ A_t \in \mathbb R^{N \times N}$ is the adjacency matrix of the $t$-th slice of TEGs, $\tilde{A_t}=A_t+I_{N(t)}$ is the adjacency matrix with self-connections. $\tilde{D}_{t(ii)}=\sum_{j} \tilde{A}_{t(ij)}$ denotes the degree matrices of $\tilde{A_t}$. $h_{t-1}$ is the evolution features of the $t$-th slice, which will be described in detail later. $W_0 \in \mathbb R^{N \times H}$ and $W_1 \in \mathbb R^{H \times d}$ denote the weight matrix of the hidden layer and the output layer, respectively. $\sigma$ is the Relu active function and the input $h_0= X$.
\setlength{\parskip}{0\baselineskip}

For the evolution process of the structural features, EF-Extractor first calculates the update gate $z_t$ and the reset gate $r_t$ according to the current structural features $Z_t$ and the previous evolution features $h_{t-1}$, which can be expressed as:
\begin{equation}
	\label{eq3}
	z_t = \sigma(Z_tW_z+h_{t-1}U_z)
\end{equation}

\setlength{\parskip}{-0.6\baselineskip}
\begin{equation}
	\label{eq4}
	r_t = \sigma(Z_tW_r+h_{t-1}U_r)
\end{equation}
where $W_z,W_r \! \in \! \mathbb R^{N \times d}$ and $U_z,U_r \!\in\! \mathbb R^{d \times d}$ are the weight matrix of the update/reset gate, respectively. The update gate decides how much $h_{t-1}$ is passed to the future, and the reset gate determines how much $h_{t-1}$ need to be forgotten.
\setlength{\parskip}{0\baselineskip}

The next step is to calculate the candidate hidden state $\tilde h_t$ by reset gate. Here, EF-Extractor stores historical evolution features $h_{t-1}$ and memorizes the current state:
\begin{equation}
	\label{eq5}
	\tilde h_t = tanh(WZ_t+(r_t \odot h_{t-1}U)
\end{equation}
where $W\in \mathbb R^{N \times d}$ and $U\in \mathbb R^{d \times d}$ are the weight matrix used to calculate $\tilde h_t$. $\odot$ denotes the Hadamard product.

\setlength{\parskip}{0\baselineskip}
Finally, EF-Extractor updates the current evolution features $h_t$ according to $h_{t-1}$ and $\tilde h_t$:
\begin{equation}
	\label{eq6}
	h_t = (1-z_t) \odot h_{t-1} + z_t \odot \tilde h_t
\end{equation}

\setlength{\parskip}{0\baselineskip}

\textbf{Evolution graphs pooling.} Intuitively, addresses with similar evolution features can be divided into the same address clusters. This motivates us to aggregate similar addresses until all addresses' evolution features are aggregated into the TEG's graph-level behavior features. The key idea of the evolution graphs pooling is to learn the cluster assignment matrix by GNNs and assign similar addresses to new address clusters. For the evolution features $h_t$ in $t$-th slice ($t\in \{1,...,T\}$), the evolution graphs pooling first calculates the current cluster assignment matrix $C_t$:
\begin{equation}
	\label{eq7}
	C_t=softmax \left ( GNN_{pool}(A_t,h_t) \right )
\end{equation}
where $GNN_{pool}$ can be any GNNs, here we choose the GCN module with the same structure as EF-Extractor. $C_t\in \mathbb R^{N\times \tilde N}$ means that $N$ addresses are assigned to $\tilde N$ new address clusters. $\tilde N = N*r$ and $r$ is the assignment ratio.

According to the adjacency matrix $\{A_1,...,A_T\}$ of the current TEG, address evolution features $\{h_1,...,h_T\}$, and the assignment matrix $\{C_1,\cdots,C_T\}$, the process of the evolution graphs pooling for $t$-th slice can be formulated as:

\setlength{\parskip}{-0.8\baselineskip}
\setlength{\abovedisplayskip}{-1.5pt}
\begin{equation}
	\label{eq8}
	h_{t}^{pool}=C_t^T h_t \in \mathbb{R}^{\tilde N \times d}
\end{equation}

\setlength{\parskip}{-1\baselineskip}
\begin{equation}
	\label{eq9}
	A_{t}^{pool}=C_t^T A_t C_t \in \mathbb{R}^{\tilde N \times \tilde N}
\end{equation}
where $d$ is the dimension of the node representation $Z$. Eq.~\ref{eq8} and Eq.~\ref{eq9} generate the evolution features $\{h_1^{pool},...,h_T^{pool}\}$ and the adjacency matrix $\{A_1^{pool},...,A_T^{pool}\}$ for $\tilde N$ address clusters, respectively. They are input to the next EF-Extractor to capture the evolution features of next address clusters.

\setlength{\parskip}{0\baselineskip}
\textbf{Behavior recognition.}  In some cases, phishers initiating malicious transactions in a specific evolution period may seriously affect the subsequent transaction evolution features. Therefore, we comprehensively consider the transaction behavior features in all slices rather than using only the most recent time slice, which can alleviate the negative impact of malicious transactions.

After extracting the evolution features $\{h_1^{pool},...,h_T^{pool}\}$  for $T$ time slices ($\tilde N =1$, $h_t \in \mathbb R^{1 \times d},t\in [1,...,T]$), the \emph{Read-out} operation assigns time coefficients $\alpha=[\alpha_1,...,\alpha_T]$ to different evolution features, aggregating the $T$ evolution features into the unique evolution feature $h_i$ for the target address $v_i$:

\setlength{\parskip}{-0.8\baselineskip}\vspace{-0.2em}
\setlength{\abovedisplayskip}{-2pt}
\begin{equation}
	\label{eq10}
	h_i = \sum_{t=1}^T \alpha_t h_t^{pool}
\end{equation}

\setlength{\parskip}{-0.5\baselineskip}
\noindent where $h_t^{pool}$ denotes $v_i$'s evolution features of $t$-th slice.
\setlength{\parskip}{0\baselineskip}

Finally, we take $h_i$ as the input of the \emph{MLP} layer with a softmax classifier. Moreover, we use the cross-entropy function $\mathcal{L} $ to train TEGDetector, which is given by:

\setlength{\parskip}{-0.4\baselineskip}
\begin{equation}
	\label{eq11}
	\hat{Y}= softmax(MLP(h_i))
\end{equation}

\setlength{\parskip}{-1.1\baselineskip}
\begin{equation}\label{eq12}
	\mathcal{L} = -\sum_{\mathcal G_i \in \mathbb G_{set}} \sum_{j=1}^{|Y|}Q_{ij}\ln{\hat{Y}_{ij}\left(A^i,X^i\right)}
\end{equation}
where $\mathcal G_i \in \mathbb G_{set}$ denotes the target address $v_i$'s TEG in the training set $\mathbb G_{set}$. $Y=\{y_1,...,y_n\}$ is the category set of the TEGs. $Q_{ij}=1$ if $\mathcal G_i$ belongs to category $y_i$ and $Q_{ij}=0$ otherwise. $\hat{Y}_{ij}$ denotes the predicted probability of $\mathcal G_i$, which is calculated by Eq.\ref{eq11} and can be considered as a function of $A^i$ and $X^i$, thus we denote it as $\hat{Y}_{ij}(A^i,X^i)$.
\setlength{\parskip}{0\baselineskip}

\section{Experiments}
In this section, we comprehensively evaluate the proposed TEGDetector, including its phishing detection performance, detection efficiency, and robustness.

%
%
%
%

\begin{table}[htbp]\setlength{\abovecaptionskip}{0.05cm}\setlength{\belowcaptionskip}{-0.2cm}\setlength{\abovecaptionskip}{0.1cm}\vspace{0.2cm}
	\centering
	\caption{Dataset statistics}
	\resizebox{80mm}{9.6mm}{
		\begin{tabular}{c|ccc}
			\hline \hline
			TEG properties   & \# Addresses  & \#Transctions  & Average degree \\ \hline
			Sum &790,849  &3,383,022  &-  \\
			Average &395.42  &1,691.51  &4.86  \\
			Maximum &4,934  &110,060  &7.09  \\
			Minimum &2  &1  &1.00  \\ \hline \hline
		\end{tabular} \label{data} }
\end{table}
\vspace{-0.4cm}

\begin{table*}[htbp]\setlength{\abovecaptionskip}{0.1cm}\setlength{\belowcaptionskip}{0.3cm}
	\centering
	\caption{The detection performance of different phishing detectors. We use bold to highlight wins.}
	{
		\begin{tabular}{cccccccccc}
			\hline \hline
			\multirow{3}{*}{Compared methods} & \multicolumn{9}{c}{Training ratio}                                                                                                                                                               \\ \cline{2-10}
			& \multicolumn{3}{c|}{60\%}                                              & \multicolumn{3}{c|}{70\%}                                             & \multicolumn{3}{c}{80\%}                        \\ \cline{2-10}
			& Precision      & Recall         & \multicolumn{1}{c|}{F-score}        & Precision      & Recall         & \multicolumn{1}{c|}{F-score}        & Precision      & Recall         & F-score        \\ \hline
			Density detector            &50.41           &\textbf{99.30 }           & \multicolumn{1}{c|}{66.87}         &50.41           &\textbf{99.30 }             & \multicolumn{1}{c|}{66.87}        &50.41           &\textbf{99.30 }              & 66.87  \\
			Repeat detector           &46.94           &51.78           & \multicolumn{1}{c|}{49.25}          &48.16           &62.35           & \multicolumn{1}{c|}{54.35}          & 48.53          &69.62           &57.19    \\
			Deepwalk                   & 76.85          & 68.85          & \multicolumn{1}{c|}{72.63}          & 77.90          & 71.20          & \multicolumn{1}{c|}{74.40}          & 78.15          & 72.70          & 75.33          \\
			Node2vec                   & 81.65          & 70.35          & \multicolumn{1}{c|}{75.58}          & 82.30          & 72.20          & \multicolumn{1}{c|}{76.92}          & 82.65          & 74.85          & 78.56          \\
			Trans2vec                  & 78.65          & 77.90          & \multicolumn{1}{c|}{78.27}          & 88.65          & 86.55          & \multicolumn{1}{c|}{87.59}          & 91.45          & 87.65          & 89.51          \\
			T-EDGE                     & 79.05          & 76.20          & \multicolumn{1}{c|}{77.60}          & 87.45          & 76.65          & \multicolumn{1}{c|}{81.69}          & 88.75          & 78.55          & 83.34          \\
			I$^2$BGNN                     & 88.65          & 91.45          & \multicolumn{1}{c|}{90.03}          & 89.20          & 91.55          & \multicolumn{1}{c|}{90.36}          & 89.20          & 92.05          & 90.60          \\
			MCGC                       & 90.50          & 91.55          & \multicolumn{1}{c|}{91.02}          & 90.55          & 92.10          & \multicolumn{1}{c|}{91.32}          & 90.75          & 92.85          & 91.79          \\
			TEGDetector (Ours)             & \textbf{95.90} & 95.60 & \multicolumn{1}{c|}{\textbf{95.75}} & \textbf{96.55} & 96.75 & \multicolumn{1}{c|}{\textbf{96.65}} & \textbf{96.30} & 96.25 & \textbf{96.28} \\ \hline \hline
		\end{tabular}\label{tab:1}}
\end{table*}

\subsection{Datasets}\label{sec4.2}
We evaluate TEGDetector on the real-world Ethereum transaction dataset released on the Xblock platform. Xblock provides 1,660 phishing addresses that have been reported and 1,700 randomly selected normal ones with the records of their two-order transactions. Specifically, we randomly selected 1,000 phishing addresses and the same number of ordinary addresses and construct the TEGs with 10 time slices ($T=10$) for them. To make a comprehensive evaluation, we divide the TEGs into two parts: \{60\%, 70\%, 80\%\} as the training set and the remaining \{40\%, 30\%, 20\%\} as the test set. The basic statistics are summarized in Table~\ref{data}. In the experiment, we repeated the above steps five times and reported the average phishing detection performance.

\subsection{Compared Methods}
To better evaluate the detection performance of TEGDetector, we choose several phishing detectors as the compared methods.
For all compared methods, we select the same target addresses as TEGDetector, and conduct experiments based on the source code released by the authors and their suggested parameter settings. The compared methods are briefly described as follows:

\textbf{Density detector} calculates the density ratio of TEG slices containing transactions to all slices. When the density ratio is greater than $0.5$, the target address will be classified as a phishing address.
\textbf{Repeat detector} calculates the repetition ratio of test transactions with the same direction as the training ones to all test transactions. According to the conclusion of Lin et al.~\cite{lin2021evolution}, we classify addresses with a repetition ratio greater than $0.1$ as phishing addresses. Note that in this case, we divide the slices in each TEG into training slices and test ones according to different division ratios.
\textbf{Deepwalk}~\cite{perozzi2014deepwalk} and \textbf{Node2vec}~\cite{grover2016node2vec} learn node representations through random walking and can be used in blockchain transaction networks.
\textbf{Trans2vec}~\cite{wu2020phishers} and \textbf{T-EDGE}~\cite{lin2020t} consider the transaction amount and timestamps of blockchain transactions on the basis of random walking, thus achieving better phishing detection performance.
\textbf{I$^2$BGNN}~\cite{shen2021identity} and \textbf{MCGC}~\cite{zhang2021mcgc} are graph classifiers designed for phishing detection on the blockchain. They have achieved satisfactory detection performance and easy to be implemented for phishing detection on new addresses.

\setlength{\parskip}{0\baselineskip}

\subsection{Performance of TEGDetector }
In this section, we discuss the phishing detection performance of TEGDetector, and analyze its detection efficiency.

\textbf{Phishing detection performance.}
Compared with other detectors in Table~\ref{tab:1}, TEGDetector achieves the state-of-the-art (SOTA) performance at different training set ratios. Specifically, although the density detector achieves a recall of 99.30\%, a precision of 50.41\% indicates that it is actually invalid to detect phishing addresses based on density ratio. Compared with density detector and repeat detector, other compared methods achieve better detection performance with automated feature learning. Unfortunately, the lack of structural or temporal information restricts them from achieving more accurate phishing detection.
In contrast to Trans2vec and T-EDGE (which focus on temporal information), I$^2$BGNN and MCGC (which pay attention to structural information), TEGDetector achieves the best detection performance. This suggests that balancing structural and temporal information can more accurately capture the target address's transaction behaviors.

\textbf{Detection efficiency.}
We further study the detection efficiency of TEGDetector. Since TEGs are essentially a series of dynamic subgraphs, we compare TEGDetector with the GNN based phishing detectors which input the static ones.



\begin{figure}[htb]\setlength{\belowcaptionskip}{-0.8cm} \setlength{\abovecaptionskip}{0.1cm}\vspace{-0.1cm}
	\centering
	\includegraphics[width=1\linewidth]{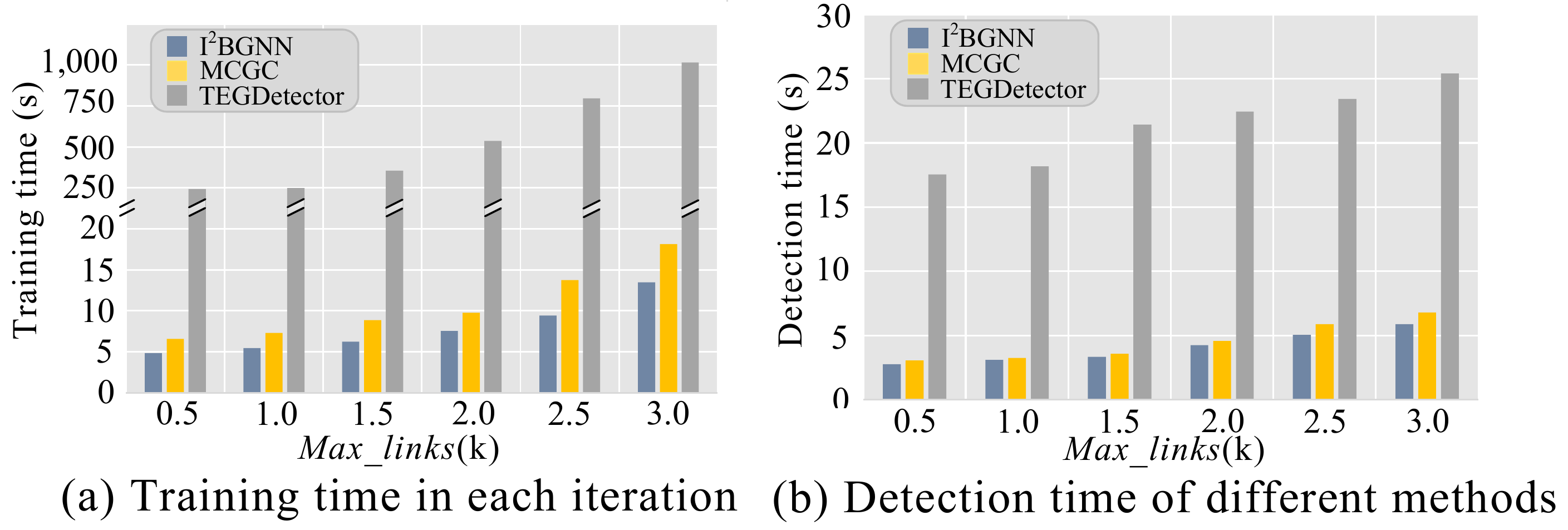}\\
	\caption{Detection efficiency of different detectors. }\label{fig6}
\end{figure}

Figure~\ref{fig6}(a) and (b) show the training time and detection time of different detectors. We select all training addresses for model training and count the detection time of 100 randomly selected addresses in the detection phase. We can observe that with the increase of $max\_\ links$, the training time of TEGDetector becomes longer than that of I$^2$BGNN. Moreover, TEGDetector's detection time still reaches 6.5 times that of I$^2$BGNN, although the gap has been greatly reduced. Considering TEGDetector's excellent detection performance, we believe that such a price is acceptable.

\subsection{A Fast and Non-parametric Phishing Detection}\label{sec4.6}
The previous section confirms the SOTA phishing detection performance of TEGDetector. However, the high time complexity of TEGDetector is still a challenge. To Address this problem, we propose to quickly filter out the obvious normal addresses while ensuring that the real phishing ones are not miss, which can narrow down the search space of suspicious addresses.


Phishers usually send phishing messages to massive users, allowing them to have more potential transaction partners. We believe that they may have more intensive large-amount transactions compared to the normal addresses. To verify our conjecture, we defined central transaction ratio (CTR), which represents the ratio of central address's transactions to all transactions in a TEG (or a static subgraph).

\begin{figure}[htb]\setlength{\belowcaptionskip}{-0.2cm} \setlength{\abovecaptionskip}{0.1cm}\vspace{0cm}
	\centering
	\includegraphics[width=1\linewidth]{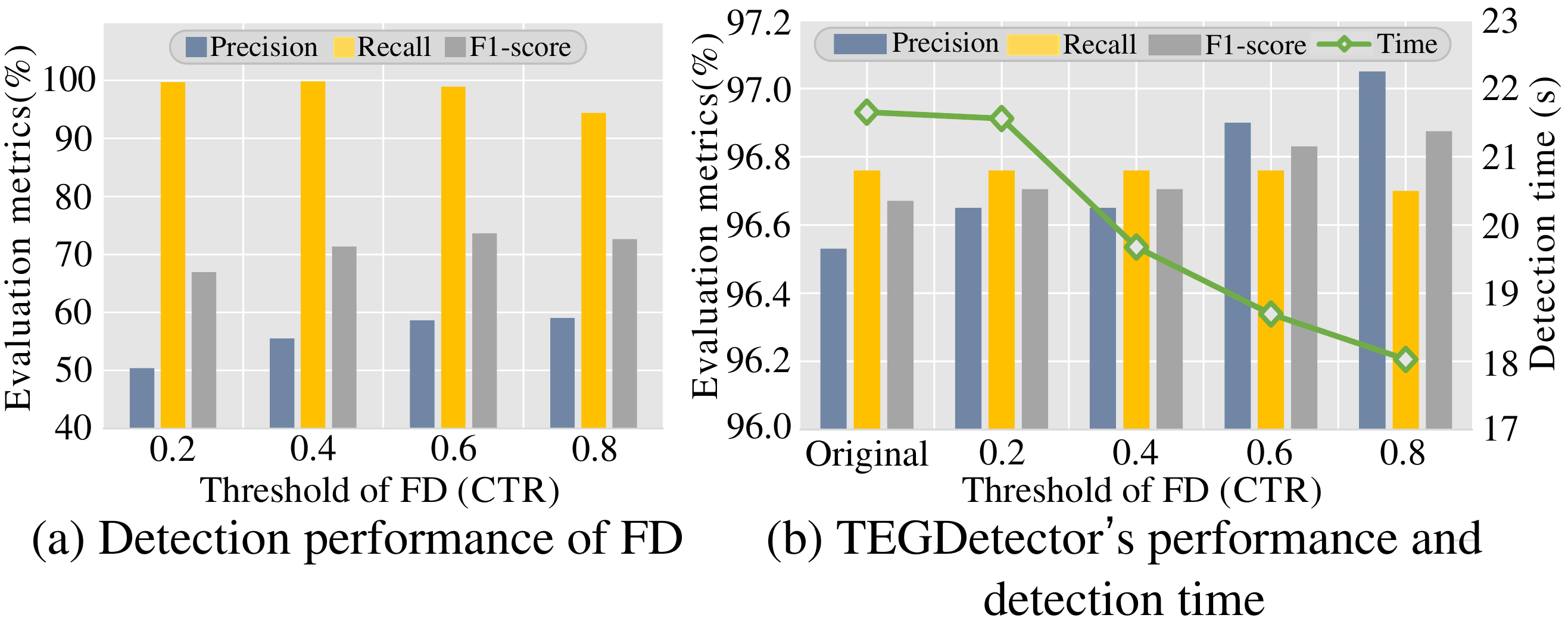}\\
	\caption{ (a) FD's detection performance under different CTR thresholds. (b) FD can improve the detection performance of TEGDetector and reduce its detection time.  }\label{fig8}
\end{figure}

Inspired by this observation, we propose a fast and non-parametric detector (FD). Specifically, we classify the target address whose CTR is greater than a threshold as a phishing address, otherwise, it is regarded as a normal address. In Figure~\ref{fig8}(a), when CTR is set to 0.6, FD can almost reach 100\% Recall, and Precision almost reaches the highest 60\%. This indicates that FD can filter out normal addresses almost without missing any phishing addresses, which can be a pre-detected approach before using TEGDetector for precise phishing detection. Meanwhile, Figure~\ref{fig8}(b) shows that the TEGDetector's Precision is improved more than its Recall since FD may pre-filters some normal addresses that may be misclassified by TEGDetector. Additionally, we can observe that FD can also reduce the detection time of TEGDetector by approximately 15\%.
Consequently, FD is a light-weighted solution for phishing detection when we consider both performance and efficiency.


\subsection{Ablation Study of TEGDetector }
To further explore the effectiveness of TEGDetector, we conduct ablation experiments on the pooling layer and time coefficient. For the pooling layer, we utilize the average pooling and maximum  pooling operations on the feature matrix of the graph, respectively, expressed as TEGD-ave and TEGD-max. For the time coefficients, we replace the weighting of time coefficients in TEGDetector with a summation operation to obtain the variant TEGDetector\_S.

\begin{table}[!htb]
	\caption{Ablation study of TEGDetector.}
	\label{tab:abl}
	\setlength{\tabcolsep}{2mm}
	\centering
\resizebox{0.5\textwidth}{!}{
	\begin{tabular}{ccccc}
		\hline \hline
	Ablation Module	          &Method   & Precision(\%)  & Recall(\%)  & F-score(\%) \\ \hline
	\multirow{2}{*}{Pool-method}
	&TEGD-ave &95.48  &95.53  &95.50  \\
	&TEGD-max &95.24  &95.25  &95.25 \\ \hline
	
	Time coefficient 		  &TEGDetector\_S &94.51  &94.56  &94.50  \\ \hline
	Proposed			      &TEGDetector &96.55  &96.75  &96.65  \\ \hline \hline
	\end{tabular}}
\end{table}

As illustrated in Table \ref{tab:abl}, the performance of TEGDetector is better than the two detection methods with average and maximum pooling, e.g, the precision of TEGDetector is 96.55\%, while the precision of TEGD-max is 95.24\%. This indicates that the pooling method using the cluster assignment matrix can extract graph-level features more effectively than the average and maximum pooling methods. We also observe that TEGDetector\_S is almost 2\% lower than TEGDetector on precision, recall and F-sorce, which demonstrates that time coefficients give different weights on different moments to improve model performance.

\subsection{Robustness of TEGDetector }
Now we evaluate the robustness of detectors when phishers maliciously conceal their phishing behaviors. From the perspective of network topology properties and detectors, we design two methods to add disturbances on the transaction networks, i.e., the CTR-based and gradient-based methods.
\begin{figure}[htbp]\setlength{\belowcaptionskip}{-0.5cm}\setlength{\abovecaptionskip}{0.4cm}
	\centering
	\includegraphics[width=1\linewidth]{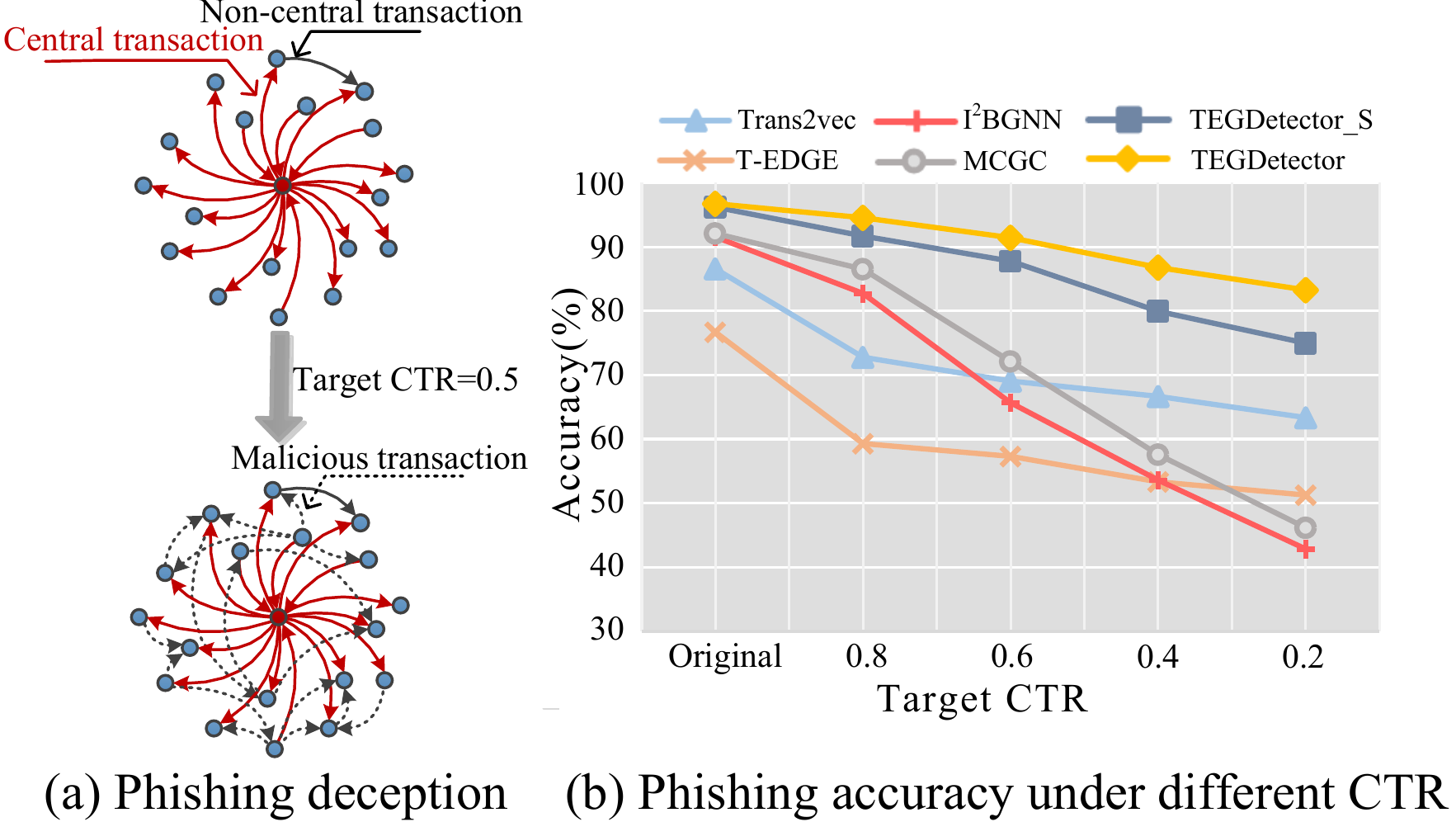}\\
	\caption{ (a) An example of phishing deception. (b) TEGDetector is more robust than other methods when facing possible phishing deception.  }\label{fig9}
\end{figure}
According to our discussion in the previous section, the phishing addresses' CTRs are more likely to be larger than 0.9. Therefore, we designed a phishing deception experiment where we change the phishing address's CTR. As shown in Figure~\ref{fig9}(a), we randomly add transactions to the non-central address pairs in TEGs until their CTRs are less than the set value. Specifically, we randomly select $T/2$ time slices in TEGs and add malicious transactions to them. The transaction amount of these transactions is set to a random value less than the maximum one in original TEGs.

In the phishing deception experiment, the smaller the target CTR, the more malicious transactions need to be added. In Figure~\ref{fig9}(b), the detection accuracy of Trans2vec and T-EDGE is more stable than other existing methods. We speculate that although the randomness of the walking strategies leads to the loss of structural information, it also reduces the impact of malicious transactions. In contrast, the static subgraphs constructed by I$^2$BGNN and MCGC retain all the malicious information, their detection accuracy reduces drastically. This further demonstrates that in addition to limiting the phishing detection performance, the robustness of static subgraphs lacking temporal information is also worrisome. Reassuringly, TEGDetector still with an accuracy of 83\% even in the worst case (when the target CTR is set to $0.2$), indicating that it is necessary to expand the static subgraphs into the dynamic ones. Compared with existing methods, TEGDetector can fully balance the target address's transaction behaviors in all periods, which enables it to capture more comprehensive behavior features. It is worth noting that TEGDetector is more robust than TEGDetector$\_$S, \emph{i.e.}, the latter undergoes an accuracy decline of 21.25\%, while the former with only a decline of 13.5\%. This testifies that TEGDetector is significantly more robust when phishers cannot add malicious transactions in each period.

\begin{figure}[htb]\setlength{\belowcaptionskip}{-0.2cm} \setlength{\abovecaptionskip}{0.1cm}\vspace{0cm}
	\centering
	\includegraphics[width=1\linewidth]{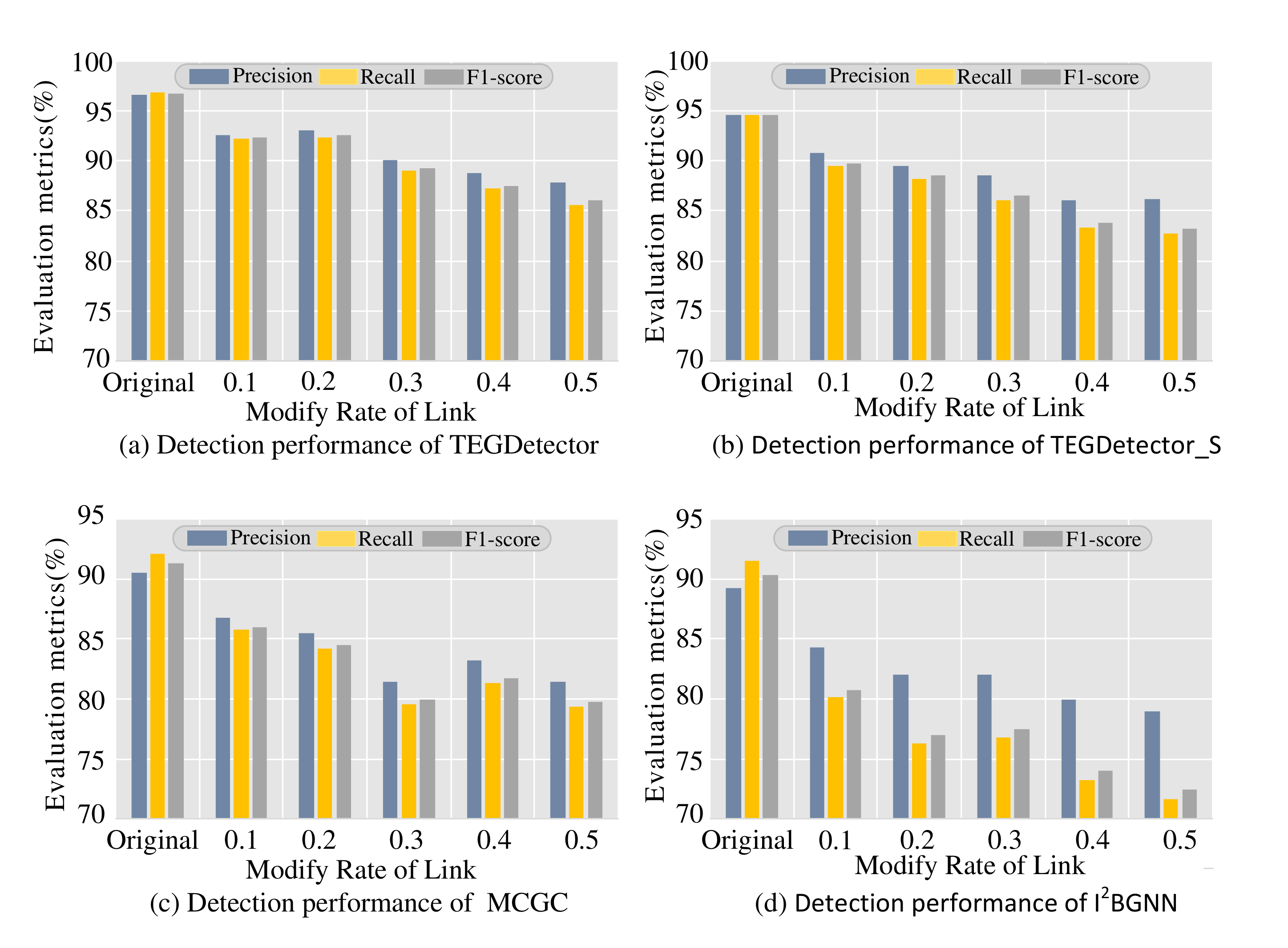}\\
	\caption{(a), (b), (c) and (d) respectively represent the performance of TEGDetector, TEGDetector\_S, MCGC, I$^2$BGNN on the perturbed transaction networks.  }\label{attack_robu}
\end{figure}

In addition to considering the malicious transactions generated based on the network topology properties, we also design the malicious transactions generated based on the feedback of the detectors gradient. Since T-EDGE and Trans2vec are based on random walk methods, these two detection methods do not perform gradient attacks and compare with other detection methods. As shown in Figure~\ref{attack_robu}, we obtain the gradient value of the transaction network adjacency matrix from the objective loss function in descending order, and add transactions in the order where there are no transactions. The value obtained by multiplying the modify rate of link by the maximum number of nodes in the transaction network is the number of malicious transactions added.

As shown in Figure~\ref{attack_robu}(a), TEGDetector exceeds 85\% in precision, recall and F1-score even if the modify rate of link is 0.5, which is better than the other three detection methods, e.g., the precision on TEGDetector\_S, the recall on MCGC and the F1-score on I$^2$BGNN are only 86.09\%, 79.33\% and 72.50\% respectively in the worst case. From the overall decline in precision,  the maximum rate of decrease in precision of TEGDetector is only 9.02\% and TEGDetector\_S is 8.91\%, while MCGC and IBGNN are 10.08\% and 11.41\% respectively. This indicates that the TEGs can disperse disturbances in multiple timestamps when mitigating the impact of malicious transactions, thereby enhancing the robustness of the detector.



\section{Conclusions}
In this paper, we first defined the transaction evolution graphs (TEGs) that can frame both structural and temporal behavior cues. Then, we proposed TEGDetector, a dynamic graph classifier suitable for identifying the target address's transaction behavior from TEGs. Experimental results demonstrate the SOTA detection performance of TEGDetector. Moreover, we gain insights that in the TEGs, large-amount transactions tend to be more concentrated on phishing addresses. Inspired by this, a fast phishing detector(FD) is designed, which can quickly narrow down the search space of suspicious addresses and improve the detection efficiency of the phishing detector. In the possible phishing deception experiment, TEGDetector shows significantly higher robustness than other phishing detectors.

However, TEGDetector's time complexity is much higher than other phishing detectors since it both considers the structural and temporal information. For future work, we plan to explore a lower complexity phishing detector. In addition, improving the robustness of phishing detectors against more targeted phishing deception methods deserves further research.

%

\bibliographystyle{IEEEtran}      
\bibliography{refer}


%

\ifCLASSOPTIONcaptionsoff
  \newpage
\fi

\end{document}